\documentclass[pra,aps,superscriptaddress,onecolumn]{revtex4} 
\usepackage{amsmath}
\usepackage{amssymb}
\usepackage{latexsym}
\usepackage{color}
\usepackage{bbm}
\usepackage{bm}

\newcommand{\dd}{\mathrm d}
\newcommand{\ii}{\mathrm i}

\def\v{\vec}

\begin{document}

\title{Einstein-Hopf drag, Doppler shift of thermal radiation and blackbody friction:\\
A unifying perspective on an intriguing physical effect}

\author{G. \L{}ach}

\affiliation{Department of Chemistry, 
University of Warsaw, Pasteura 1, 02093 Warsaw, Poland}

\affiliation{Physikalisches Institut der Universit\"{a}t, 
Albert-Ueberle-Strasse 3-5, 69120 Heidelberg, Germany}

\author{M. DeKieviet}

\affiliation{Physikalisches Institut der Universit\"{a}t, 
Albert-Ueberle-Strasse 3-5, 69120 Heidelberg, Germany}

\author{U. D. Jentschura}

\affiliation{Department of Physics, Missouri University of Science and Technology,
Rolla, Missouri 65409-0640, USA}

\affiliation{Institut f\"ur Theoretische Physik,
Universit\"{a}t Heidelberg,
Philosophenweg 16, 69120 Heidelberg, Germany}

\begin{abstract}
The thermal friction force acting on an atom moving relative to a
thermal photon bath  has recently been calculated on the basis of the
fluctuation-dissipation theorem. The thermal {\em fluctuations} of the
electromagnetic field give rise to a drag force on an atom provided one allows
for {\em dissipation} of the field energy via spontaneous emission.  The drag
force exists if the atomic polarizability has a nonvanishing imaginary part.
Here, we explore alternative derivations.  The damping of the motion of a
simple harmonic oscillator is described by radiative reaction theory (result of
Einstein and Hopf), taking into account the known stochastic fluctuations of
the electromagnetic field.  Describing the excitations of the atom as an
ensemble of damped harmonic oscillators, we identify the previously found
expressions as generalizations of the Einstein-Hopf result. In addition, we
present a simple explanation for blackbody friction in terms of a Doppler shift
of the thermal radiation in the inertial frame of the moving atom:
The atom absorbs blue-shifted photons from the front and 
radiates off energy in all directions, thereby losing energy.  The
original plus the two alternative derivations provide for additional
confirmation of an intriguing physical effect, and leave no doubt
regarding its existence.
\end{abstract}


\maketitle

%
%
\section{Introduction}

Initially, one might conjecture that the influence of incoming electromagnetic
radiation on an atom or ion should be restricted to the well-known conservative
radiation pressure force, and not involve a drag or friction force.  When
averaging the radiation pressure force of incoming electromagnetic radiation on
an atom over all directions, the effect should vanish for spatially uniform
blackbody radiation.  However, on second thought, the existence of such a drag
force is plausible.  The reason is that the fluctuations of the electromagnetic
field naturally lead to fluctuations of the atomic dipole moment, which in turn
lead to the dissipation of energy provided the atom undergoes spontaneous
emission of radiation.  According to Ref.~\cite{MkPaPoSa2003}, the friction
coefficient can be calculated based on a Green--Kubo formula, which amounts to
an integration over all possible points in times at which the dissipative
process may occur. 

As a consequence, a particle moving through a thermal radiation field
experiences this dissipation as a drag force.  The drag force acting on an atom
flying through space with a velocity $v$ with respect to the cosmic microwave
background (CMB) rest frame is nonvanishing. Already in 1910, soon after the
discovery of Planck's formula for the thermal distribution of light quanta
emerging from a perfect black body, this friction force  was calculated by
Einstein and Hopf~\cite{EiHo1910b}, for the simplified case of 
a single harmonic oscillator subject to radiative damping.
In the simple model considered by Einstein and Hopf, the Abraham--Lorentz
radiation reaction force determines the ``width'' term for the damped harmonic
motion.  Because an atom can be viewed as an ensemble of damped harmonic
oscillators, it is natural to look for possible connections of the historic
result presented by Einstein and Hopf with the recent derivation 
presented in Ref.~\cite{MkPaPoSa2003}.

A totally different point of view is reached if we 
interpret the direction-dependent drag force in terms 
of a direction-dependent Doppler effect: 
The atom sees blue-shifted, incoming photons from the front, 
whereas incoming photons from the back are red-shifted, 
while the atom radiates off the photons in all possible directions.
This leads to a drain on the energy of the atom,
which corresponds to a friction force.
The Doppler shifted, direction-dependent temperature
of blackbody radiation has been calculated in Ref.~\cite{MK1979bb}.
Incident radiation exerts a force on an atom, 
proportional to the imaginary part of the atomic 
polarizability~\cite{GrWeOv2000}.
The precise connection of these observations with the result derived 
in Ref.~\cite{MkPaPoSa2003} is clarified in the following.
The two alternative derivations of the blackbody friction force
sketched above confirm the existence
of an effect whose physical foundation has 
been called into question~\cite{MNFa2004,MkPaPoSa2004}.

SI mksA units are used throughout the article.
We proceed as follows.
The connection to the Einstein--Hopf drag is 
explored in Sec.~\ref{EH}, while an alternative derivation
based on the Doppler effect of the thermal radiation is
given in Sec.~\ref{DP}. Conclusions are drawn in Sec.~\ref{conclu}.

%
%
\section{Einstein--Hopf Drag}
\label{EH}

Roughly a century ago, Einstein and Hopf~\cite{EiHo1910b}
considered a classical charged particle, having charge $e$ and mass $m$, moving
in a one dimensional harmonic potential, and being in thermal equilibrium with
the electromagnetic field.  The equation of motion for this particle, including
the Abraham-Lorentz radiative reaction force
and the coupling to the electric field $\mathcal{E}_x$, is given by:
\begin{equation}
\label{eh}
\frac{\dd^2x}{\dd t^2} + \omega_0^2\,x - 
\sigma \frac{\dd^3x}{\dd t^3}=\frac{e}{m} \mathcal{E}_x(t)\,,
\end{equation}
with $\sigma=e^2/(6\pi\epsilon_0\,mc^3)$.  
The condition of thermal equilibrium between the atom and the thermal bath
implies a nonvanishing Einstein--Hopf (EH) 
damping force $F_{\rm EH}$ if the center of mass of the harmonic
potential is in motion with respect to the field.  In the limit of slow 
motion, the velocity-dependent force is equal to
\begin{equation}
F_{\rm EH}= -\frac{4\pi^2e^2\,v}{5m c^2 \, (4\pi\epsilon_0)}
\left( \rho(\omega_0;T) - \frac{\omega_0}{3} 
\frac{\partial \rho(\omega_0;T)}{\partial \omega_0} \right) \,,
\end{equation}
a result which has been recorded in Ref.~\cite{EiHo1910b} and in
Eq.~(4.1) in Ref.~\cite{Mi1980}.  Herein,
$\rho(\omega;T)$ represents the energy density of the electromagnetic radiation,
\begin{equation}
\rho(\omega;T) =
\frac{\hbar\,\omega^3}{\pi^2\,c^3}\,\eta(\omega;T)\,.
\end{equation}
We now use the identity
\begin{equation}
\rho(\omega;T) - \frac{\omega}{3} \frac{\partial \rho(\omega;T)}{\partial \omega}=
-\frac{\hbar\,\omega^4}{3\pi^2\,c^3}\frac{\partial\eta(\omega;T)}{\partial\omega}
\,, 
\qquad
\qquad
\eta(\omega;T)=\frac{1}{\exp\left( \beta \hbar \omega \right)-1} \,,
\end{equation}
using the thermal photon occupation number $\eta(\omega;T)$.
So,
\begin{equation}
\label{force1}
F_{\rm EH}
= \frac{4 \hbar \, e^2 \, \omega_0^4 \,v}%
{15 \, m c^2 \, (4\pi\epsilon_0)}
\frac{\partial\eta(\omega_0;T)}{\partial\omega_0} 
= \frac{ \hbar \, e^2 \, \omega_0^4 \,v}%
{15 \, \pi \, m c^2 \, \epsilon_0}
\frac{\partial\eta(\omega_0;T)}{\partial\omega_0} \,.
\end{equation}
The connection to Ref.~\cite{MkPaPoSa2003} can now be found as follows.
According to Refs.~\cite{RyKrTa1989vol3,PiLi1958vol9}, 
the polarizability tensor is isotropic for atoms, and
the main result in Ref.~\cite{MkPaPoSa2003} is an expression for the effective
friction (EF) force as the spectral integral of the imaginary part of the
dynamic dipole polarizability of the particle [see Eq.~(12) in
Ref.~\cite{MkPaPoSa2003}]:
\begin{equation}
\label{force2}
F_{\rm EF} = -\frac{\beta\hbar^2\,v}{3\pi\,c^5\,(4 \pi \epsilon_0)}
\int_0^{\infty}\dd\omega\,
\frac{\omega^5\, {\rm Im}\,\alpha(\omega)}
{\sinh^2(\tfrac12 \beta \hbar \omega)}
= \frac{\hbar\,v}{3\pi^2\,c^5\,\epsilon_0}
\int_0^{\infty}\dd\omega\,\omega^5\,
\frac{\partial\eta(\omega;T)}{\partial\omega}\,{\rm Im}\,\alpha(\omega)\,,
\end{equation}
Here, $\beta = 1/(k_B T)$ is the Boltzmann factor and $\alpha(\omega)$
represents the dynamic polarizability.  Note that $F_{\rm EF}$ points in a
direction opposite to the velocity $v$ of the particle and thus acts as a drag
force; the sign of the derivative $\partial\eta(\omega;T)/\partial\omega$ is negative.  
We have used the identity
\begin{equation}
\frac{\beta \hbar}{\sinh^2(\tfrac12 \beta \hbar \omega)}=
- 4 \frac{\partial\eta(\omega;T) }{\partial\omega}\,.
\end{equation}

In order to compare the results in Eqs.~\eqref{force1} and Eq.~\eqref{force2},
we follow an argument presented in Ref.~\cite{Mi1980}: For
randomly oriented, classical oscillators, the formula~\eqref{force2} needs to
be averaged over the projection of the electromagnetic wave vector $\v{k}$ onto
the $z$-axis (along which the potential center is moving), and over the
projections of the polarization vectors $\v{\epsilon}_{\v{k}\lambda}$ onto the
$x$-axis (the direction of harmonic oscillation), 
and over the photon polarizations.  This gives an additional
factor of
\begin{equation}
\label{215}
\frac12 \, \int \frac{\dd\Omega_k}{4\pi}\hat{k}_z^2
\sum_{\lambda}\epsilon^2_{\v{k}\lambda x}=\frac{2}{15}\ .
\end{equation}

The forces calculated using Eqs.~\eqref{force1} and~\eqref{force2} can be
compared by solving the classical equation of motion for the Einstein-Hopf
model with an oscillatory electric field. The corresponding classical dynamic
polarizability reads as follows ($\rm {HO}$ stands for a 
harmonic oscillator)
\begin{equation}
\alpha_{\rm{HO}}(\omega) = \frac{e^2}{m}
\, \frac{1}{\omega^2 - \omega_0^2 - 
\ii\,\sigma\,\omega^3} \,,
\end{equation}
which is proportional to the Green function for the 
oscillator described by Eq.~\eqref{eh}.
In the limit of small radiative damping, 
i.e.  $\sigma\ll 1/\omega_0$, the nonvanishing 
values of $\alpha_{\rm{HO}}(\omega)$ is concentrated 
in the area $\omega \approx \omega_0$, and we can approximate
\begin{equation}
{\rm Im}\,\alpha_{\rm{HO}}(\omega)
\approx \frac{\pi e^2}{2\omega_0\,m}\delta(\omega-\omega_0)\,.
\end{equation}
The matching of the atomic polarizability against that 
of a harmonic oscillator now proceeds as follows,
\begin{equation}
{\rm Im}\,\alpha(\omega)=
\frac{2}{15} \times 3 \times 
{\rm Im}\,\alpha_{\rm{HO}}(\omega)=
\frac{\pi \, e^2}{5 \, m \, \omega_0} \,
\delta(\omega-\omega_0)\,.
\end{equation}
The factor $2/15$ stems from Eq.~\eqref{215}.
The factor $3$ is due to the fact that the atomic polarizability 
(for the ground state) is isotropic, and therefore 
the factor $3$ counts the possible excitations in all three
directions of space.
The identification of $F_{\rm EF}$ as $F_{\rm EH}$ is now done 
easily,
\begin{align}
\label{force3}
F_{\rm EF} = & \;
\frac{\hbar\,v}{3\pi^2\,c^5\,\epsilon_0}
\int_0^{\infty}\dd\omega\,\omega^5\,
\frac{\partial\eta(\omega;T)}{\partial\omega}\,{\rm Im}\,\alpha(\omega)
\to
\frac{\hbar\,v}{3\pi^2\,c^5\,\epsilon_0}
\int_0^{\infty}\dd\omega\,\omega^5\,
\frac{\partial\eta(\omega;T)}{\partial\omega}\,
\frac{\pi \, e^2}{5 \, m \, \omega_0} \,
\delta(\omega-\omega_0)
\nonumber\\[2ex]
\to & \; \frac{\hbar\,e^ 2\, \omega_0^4 \, v}{15 \, \pi\,m \, c^5\,\epsilon_0}
\frac{\partial\eta(\omega_0;T)}{\partial\omega_0} = 
F_{\rm EH}\,.
\end{align}
This shows that the Einstein--Hopf and the effective-damping force
have the same physical origin and describe the same 
physical effect.

%
%
\section{Doppler Effect and Blackbody Friction}
\label{DP}

In order to derive the blackbody friction force on an atom 
based on the direction-dependent Doppler effect,
we need (i)~a result which allows us to express the shift in 
the incoming blackbody radiation frequencies in terms 
of a direction-dependent temperature,
and (ii)~a result which connects the incoming light intensity
to the imaginary (dissipative) part of the 
atomic polarizability.

The second of these prerequisites is given in Ref.~\cite{MK1979bb}.
Indeed, the combined
effects of the Doppler shift and the change in intensity due to Lorentz
contraction~\cite{MK1979bb}, visible for a moving observer, leads to a change
of the observed temperature $T_\textrm{o}$ of the thermal bath, which depends
on the relative angle $\theta$ between the wave vector and the velocity $\v{v}$
and the magnitude of the latter:
\begin{equation}
\label{To}
T_\textrm{o}(\theta) =T\,\sqrt{\frac{c-v\cos{\theta}}{c+v\cos{\theta}}}
=T-T\frac{v}{c}\cos{\theta}+\mathcal{O}(v^2)\, .
\end{equation}
This result is based on the
known properties of blackbody radiation under Lorentz
transformation~\cite{MK1979bb}. 

The first prerequisite can be found in Ref.~\cite{GrWeOv2000}.
Namely, the force acting on a particle in the field of plane
monochromatic radiation of intensity $I(\omega)$ is equal to
\begin{equation}
\label{FI}
F = -\frac{\omega}{\epsilon_0\,c}\,{\rm Im}\,\alpha(\omega)\,I(\omega)\ .
\end{equation}
In the case of a particle moving through blackbody radiation, the 
direction-dependent intensity
distribution is 
\begin{equation}
I(\omega)=\rho(\omega;T_\textrm{o}(\theta))/(4\pi) \,.
\end{equation}

The rest of the calculation is rather straightforward.
A particle moving through blackbody radiation experiences a
nonzero force parallel to its velocity.
In view of the above, we have
\begin{equation}
\begin{split}
F_\parallel \sim
\int_0^\infty\dd\omega \, 
\left( -\frac{\omega}{\epsilon_0\,c} \, 
{\rm Im}\,\alpha(\omega) \right)\,
\int\dd\Omega\,\cos{\theta}\,
\frac{\rho\big(\omega;T_\textrm{o}(\theta)\big)}{4\pi}\,.
\end{split}
\end{equation}
Using the chain rule, we have
\begin{equation}
\rho\big(\omega; T_\textrm{o}(\theta)\big) =
\rho\big(\omega;T - T \, \frac{v}{c} \cos{\theta} \big) \approx
\rho\big(\omega;T) + \frac{\partial}{\partial T} \rho\left(\omega;T\right) \,
\left( - T \, \frac{v}{c} \cos{\theta} \right) \,.
\end{equation}
Concerning the sign of the force, we note that for positive 
driving frequencies, the imaginary part of the atomic 
polarizability in Eq.~\eqref{FI} is positive.
The force $F$ given in Eq.~\eqref{FI} is 
thus negative and slows the atom.
The derivative $\partial \rho\left(\omega;T\right)/
\partial T$ is positive. We know that the effective 
temperature of blackbody radiation has to increase 
in the flight direction of the atom, i.e., when $\theta = 0$.
By contrast, in formula~\eqref{To}, $\theta$ measures the 
angle with respect to the incoming radiation, not the 
angle with respect to the counterpropagating direction.
We thus insert a minus sign 
and extract the velocity-dependent part of Eq.~\eqref{To} as follows,
\begin{align}
F_\parallel =& \;
\int_0^\infty\dd\omega \, 
\left( -\frac{\omega}{\epsilon_0\,c} \, 
{\rm Im}\,\alpha(\omega) \right)\,
\int \frac{\dd\Omega}{4\pi}\,\cos{\theta}\,
\left( \frac{\partial}{\partial T} \rho\left(\omega;T\right) \right) \,
\left( + T \, \frac{v}{c} \cos{\theta} \right)
\nonumber\\[2ex]
=& \;
-\frac{v}{3\epsilon_0\,c^2}
\int_0^\infty\dd\omega\,\omega\,T\,
\frac{\partial \rho(\omega;T)}{\partial T}\,{\rm Im}\,\alpha(\omega)
\nonumber\\[2ex]
=& \;
-\frac{v}{3\epsilon_0\,c^2}
\int_0^\infty\dd\omega\,\omega\,
\left( -\frac{\hbar\,\omega^4}{\pi^2\,c^3}
\,\frac{\partial \eta(\omega;T)}{\partial \omega} \right) \,
\,{\rm Im}\,\alpha(\omega)
\nonumber\\[2ex]
=& \;
\frac{\hbar v}{3 \, \pi^2 \,c^5\, \epsilon_0}
\int_0^\infty\dd\omega\,\omega^5\,
\frac{\partial \eta(\omega;T)}{\partial \omega} \,
\,{\rm Im}\,\alpha(\omega) \,,
\end{align}
which is exactly equal to Eq.~\eqref{force2}.
We have used the identity 
\begin{equation}
T\frac{\partial \rho(\omega;T)}{\partial T}=-\frac{\hbar\,\omega^4}{\pi^2\,c^3}
\,\frac{\partial \eta(\omega;T)}{\partial \omega}\,,
\end{equation}
which is elementary.
We are thus able to 
invoke a simple picture, in which the drag introduced in~\eqref{force2} can
ultimately be traced back to the Doppler effect.  A priori, the stochastic
field might be thought to increase the kinetic energy of the classical
particle, because the fluctuations continuously kick the particle.  Once the
particle is in motion, however, the thermal equilibrium condition leads to a
decrease in the kinetic energy down to the recoil limit.  Thermal photons
approaching the moving particle from the front are blue-shifted, whereas
photons coming in from the back are red-shifted and thus less energetic.  In
contrast, the emission is symmetric with respect to the forward and backward
directions in the rest frame of atom.  At thermal equilibrium between the
moving particle and the fluctuating field we thus have a net dissipation of
energy, experienced as drag, or blackbody friction.

%
%
\section{Conclusions}
\label{conclu}

We provide two alternative
derivations for the blackbody friction force on a 
moving atom, originally derived in Ref.~\cite{MkPaPoSa2003}
but called into question~\cite{MNFa2004,MkPaPoSa2004}.
In particular, we show that the Einstein--Hopf
result for thermal friction can be interpreted in terms of the 
thermal friction force on an atom, provided the atomic polarizability tensor is
replaced by the one corresponding to a classical object under the influence of
a damping force, that is proportional to the time derivative of the
acceleration (radiative reaction). 

We also present an intuitive picture, in which the blackbody
friction force is ascribed to a change in temperature of the thermal bath, as
observed by an atom moving through it, for the thermal photons coming from
different directions.  Indeed, the {\em combined} effect of the Lorentz
contraction and the corresponding change in intensity of the blackbody
radiation and of the Doppler shift can be absorbed in a direction-dependent
temperature change for the thermal photons.  Expanding the temperature change
to first order in the velocity, we give an immediate and straightforward
rederivation of the effect. 

We hope that these considerations not 
only lead to a confirmation of the expressions used earlier
for the evaluation of black-body friction~\cite{MkPaPoSa2003}, 
but also to a deeper understanding of this phenomenon.
Last, but not least, let us mention that the precise 
numerical evaluation of the integral expression 
given in Eq.~\eqref{force1} leads to very interesting problems,
both on the conceptual as well as on the numerical side.
These are discussed in Ref.~\cite{LaDKJe2011prl}.
Indeed, it turns out that the numerical evaluation of the 
integral
\begin{equation}
F_{\rm EF} = -\frac{\beta\hbar^2\,v}{3\pi\,c^5\,(4 \pi \epsilon_0)}
\int_0^{\infty}\dd\omega\,
\frac{\omega^5\, {\rm Im}\,\alpha(\omega)}
{\sinh^2(\tfrac12 \beta \hbar \omega)} \,,
\end{equation}
is highly nontrivial for medium and small temperatures:
The thermal factor (the hyperbolic sine) in the denominator
drops so fast with the frequency that the tail of the 
Lorentzian line curves in the polarizability become important.
This leads to a very peculiar behaviour of the 
friction force as a function of the temperature,
as outlined in Ref.~\cite{LaDKJe2011prl}.

%
%
\section*{Acknowledgments}

This project was supported by the National Science
Foundation and by a precision measurement grant from the National Institute of
Standards and Technology.  G.L.~acknowledges support by the Deutsche
Forschungsgemeinschaft (DFG, contract Je285/5--1).

\end{document}